# Hydrogen bond relaxation dynamics and the associated symmetric, volumetric, vibronic, and phase transitional anomalies of frozen $H_2O$ under compression

Chang Q Sun[1,2] Xi Zhang,[1] Weitao Zheng[3]

[1] *School of Electrical and Electronic Engineering, Nanyang Technological University, Singapore 6397982*

[2] *Faculty of Materials and Optoelectronic Physics, Xiangtan University, Hunan 411105, China*

[3] *Department of materials Science, Jilin University, Changchun Changchun 130012, China*

*E-mail:* Ecqsun@ntu.edu.sg

Abstract

Coulomb repulsion between the unevenly-bounded bonding "-" and nonbonding ":" electron pairs in the "$O^{2-}$ : $H^{+/p}$-$O^{2-}$" hydrogen-bond is found to originate the anomalies of low-compressibility, phonon relaxation dynamics, proton symmetrization in the hydrogen-bond, and the change of the critical temperature for the VIII-VII phase transition of ice under compression. The resultant force of the compression, the repulsion, and the uneven binding strength of the electron pairs make the softer intermolecular "$O^{2-}$ : $H^{+/p}$" nonbonding lone pair be highly compressed and stiffened but the stiffer intramolecular "$H^{+/p}$-$O^{2-}$" bond be elongated and softened. Consequently, the softer nonbond phonons (< 400 $cm^{-1}$) are stiffened and the stiffer bond phonons (> 3000 $cm^{-1}$) are softened upon compression. The nonbond compression and the real bond elongation results in the $O^{2-}$-$H^{+/p}$ : $O^{2-}$ symmetrization and the low compressibility of ice. Findings should form the starting point to unveil the physical anomalies of $H_2O$ under various stimuli.

I       Introduction

$H_2O$ has been the subject of extensive study, given its paramount importance in the nature science[1, 2, 3, 4, 5] and its key role in DNA folding,[6, 7] protein and gene delivery.[8, 9] Although the structure formation and the phase transition under compressing or cooling [10, 11, 12] and the reaction dynamics[13, 14] of $H_2O$ with other ingredients have been intensively investigated, many fundamental questions about the properties of water away from the ambient conditions remain yet puzzling.[15, 16, 17, 18, 19, 20] For instances, it is usual for other materials that the crucial temperature ($T_C$) for liquid-solid or disordered-ordered phase transition increases with the pressure ($P$) in a quasi-equilibrium process;[21, 22] however, the $T_C$ for Ice-VIII phase transferring to the proton-disordered Ice-VII phase drops from 280 to 150 K when the P is increased from 1 to 50 GPa.[23] Under compression, the O---O distance becomes shorter while the O-H bond length becomes longer, leading to the low-compressibility and the proton symmetrization of Ice-VIII at 59 GPa and at a O---O distance of 0.20 nm.[5, 24] Generally, the applied pressure stiffens all the Raman phonons of other materials such as carbon allotropes;[25] however, the ice-VIII vibration spectra demonstrated the anomalous stiffening of the softer mode at frequency lower than 400 cm$^{-1}$ but softening of the stiffer mode at frequency greater than 3000 cm$^{-1}$. [4] Such vibronic anomaly has also been observed by other Raman [23] and infrared (IR) absorption[26] measurements. Unfortunately, correlation between the hydrogen bond and the phonon relaxation dynamics as well as the observed anomalies of frozen $H_2O$ under compression has been far from clear. These anomalies are beyond the expectation of Pauling's Ice Rule[27] that predicted the crystal geometry and entropy of various phases.

The aim of this communication is to show that a combination of the extended Ice Rule of Pauling[27] and the *ab initio* molecular dynamics (MD) calculations has enabled us to clarify and correlate these concerns leading to comprehension of the origin of the symmetric, vibronic, volumetric and $T_C$ anomalies of frozen $H_2O$ under compression from the perspective of the relaxation dynamics of the real and the virtual bond segments in the hydrogen bond.

II      Hypothesis: the extension of Pauling's Ice Rule

Figure 1(a) illustrates the extension of Pauling's ice rule[27] (the central tetrahedron) and (b) the segmentation of one "$O^{2-}$ : $H^{+/p}$-$O^{2-}$" hydrogen bond showing the principle of pressure-induced proton symmetrization of the hydrogen bond. This extension contains two $H_2O$ molecules and four identical quasi-linear "$O^{2-}$ : $H^{+/p}$-$O^{2-}$" hydrogen bonds.[28] The $H^{+/p}$ plays a dual role of $H^+$ and $H^P$ as it donates its electron to one $O^{2-}$ and meanwhile it is polarized by the nonbonding lone pair of the other neighboring $O^{2-}$ upon the sp-orbit of oxygen being hybridized in reaction.[29] In the hexagonal or cubic ice the $O^{2-}\cdots O^{2-}$ distance is 0.276 nm. The intramolecular $H^{+/p}$-$O^{2-}$ bond is much shorter and stronger (~0.100 nm and ~$10^0$ eV) than that of the intermolecular $O^{2-}$ : $H^{+/p}$ nonbond (~0.176 nm and ~$10^{-2}$ eV). The angle between the $H^{+/p}$–$O^{2-}$–$H^{+/p}$ is smaller than 104.5° while the angle between the $H^{+/p}$ : $O^{2-}$ : $H^{+/p}$ is greater than 108.5° for a free $H_2O$ molecule.

The advantage of the extended Ice Rule is that it allows us to focus on the two segments of the nonbonding lone pair ":" virtual bond and the bonding pair "–" of the real bond in one of the four "$O^{2-}$ : $H^{+/p}$-$O^{2-}$" bonds. This segmentation allows us to examine the responses of the intramolecular nonbonding lone pair ":" and the intermolecular bonding pair "–" of electrons to the applied stimulus and their correlation, for example, the proton symmetrization and the frequencies of vibration of the two segments under pressure, as shown in Figure 1(b). The $H^{+/p}$ was taken as the coordinate origin.

We hypothesized that the $O^{2-}$ : $H^{+/p}$ lone pair is readily to be compressed. The combination of the compression, the Coulomb repulsion between the unevenly-bounded ":" and "-" electron pairs and their respective binding strength determines the displacements of the electron pairs and hence the lengths and strengths of the real and the virtual bond. If the $O^{2-}$ : $H^{+/p}$ is compressed, the ":" charge center moves towards the $H^{+/p}$ and then the $H^{+/p}$-$O^{2-}$ bonding pair is pushed slightly away from the $H^{+/p}$ origin. The $H^{+/p}$-$O^{2-}$ bond becomes longer and weaker but the result $O^{2-}\cdots O^{2-}$ will be shorter. In order to verify the hypothesis, we conducted the *ab initio* molecular dynamics (MD) calculations and Raman spectroscopic analysis to examine the length and stiffness correlation between the two electron pairs.

III    MD calculations

The MD calculations were performed using Forcite package with *ab initio* optimized force field Compass27.[30] The evolution of the O-H and O:H distances in a unit cell containing 32- molecules of ice-VIII under the pressure changing from 1 to 20 GPa was dynamically relaxed under constant pressure and constant enthalpy (NPH) ensemble for 30 ps. The average O-H and O:H lengths were taken of the structures of the last 10 ps (20,000 steps). The power spectra were calculated from the Fourier transform of their velocity autocorrelation function using velocity data of all atoms recorded at every 0.5 fs. The Fourier transform of the velocity autocorrelation function can be expressed as:[31]

$$I(\omega) = F(Cor(t)) = 2\int_0^\infty Cor(t)\cos\omega t\, dt$$

$$\text{With } Cor(t) = v(\tau) * v(-\tau) = \frac{\sum_{j=1}^{n_t}\sum_{i=1}^{N} v_i(\tau_j+t)v_i(\tau_j)}{\sum_{j=1}^{n_t}\sum_{i=1}^{N} v_i^2(\tau_j)}$$

Where $v_i$ is the velocity of the $i$th atom and * represents convolution. The velocities were taken from the last 10 ps (20,000 steps).

IV    Results and discussion

4.1 Proton central symmetrization and the compressibility

Table 1 lists and Figure 2(a) shows the relaxation dynamics of the bonding and nonbonding segments as a function of pressure. When the pressure is increased from 1 to 20 GPa the $H^{+/p} : O^{2-}$ nonbond is compressed from 0.1767 to 0.1692 nm and meanwhile the $H^{+/p}$-$O^{2-}$ bond is elongated from 0.0974 to 0.1003 nm, with the polynomial expressions of the P-dependent length, $d_x = d_0[1+\alpha(P-P_0)+\beta(P-P_0)^2]$. The calculated O-H and O---O length changing trends agree with those reported in refs [5, 32, 33]. Table 1 shows the $P_0 = 1$ GPa and the derived $d_0$, $\alpha$, $\beta$ values.

It is exciting, as shown in Figure 2(a), that the extrapolation of the polynomial expressions for the two segments leads to the proton symmetrization occurring at 58.6 GPa with the $O^{2-}$---$O^{2-}$ distance of 0.221 nm, which is in exceedingly good accordance with the reported results of 59 GPa and 0.220 nm.[24] Holzapfel[34] predicted 40 years ago that, under pressure, hydrogen bonds might be transformed from the highly asymmetric $O^{2-}$-$H^{+/p} : O^{2-}$ configuration to a symmetric state in which the $H^+$ proton lies midway between the two $O^{2-}$, leading to a non-molecular symmetric phase of ice, as numerically confirmed by Benoit, Marx, and Parrinello in 1998 with the proposed mechanism of "translational proton quantum tunneling under compression".[24] Figure 2(b) compares the MD calculated and the measured[4] pressure-dependent volume of ice-VIII, with the fitting equation as $V/V_0 = 1 - 2.38\times10^{-2}P + 4.70\times10^{-4}P^2$ and $V_0$=1.06 cm$^3$/kg in experiment and 1.02 cm$^3$/kg in calculation. The *in situ* high-pressure and low-temperature synchrotron x-ray diffraction and optical Raman spectroscopy were conducted by Yoshimura et al.[4]

Consistency between the MD derived proton symmetrization and volume compression and the experimental results evidence the validity of the hypothesis that the weaker lone pair is highly compressive yet the stronger bonding pair is elongated by the compression and the repulsion. Therefore, the current MD derivatives may represent the true situations of the H-bond relaxation dynamics of frozen $H_2O$ under compression.

4.2 The relative force constant for the bond and nonbond

It is clear that the weaker $O^{2-}:H^{+/p}$ nonbond is highly compressed but the $H^{+/p}$-$O^{2-}$ bond is slightly elongated towards the proton centralization at a certain pressure because of the repulsion between the unevenly-bounded electron pairs. According to findings, we can estimate the relative force constant of each segment in the hydrogen bond. In addition to the compression force $F_p$ and the repulsion force $F_q$ the deformation forces of $k_H \delta d_H$ and $k_L \delta d_L$ opposing to the dislocation directions determines the equilibrium of the electron pairs in the hydrogen bond. The $k_H$ and $k_L$ are the force constant and $\delta d_H$ and $\delta d_L$ are the dislocation of the respective electron pairs. The equilibrium of these three forces acting on each electron pair leads to,

$$k_H \delta d_H + k_L \delta d_L = 0, \text{ or, } k_H(\delta d_H/\delta p) + k_L(\delta d_L/\delta p) = 0,$$

$$\frac{k_H}{k_L} = -\frac{\delta d_L/\delta p}{\delta d_H/\delta p}$$

This expression indicates that the force constant of each segment is proportional to the inverse slope of the respective $d_H$-P and the $d_L$-P curves. The slopes are opposite in sigh: the nonbond bends down and the real bond up, being the case of Figure 2(a). The resultant of the two slopes explains why the compressibility of ice is anomalously low.

4.3 Raman phonon relaxation

Figure 3(a) shows the MD derived pressure dependence of the power spectra in the specified frequency ranges of $\omega_L < 400$ cm$^{-1}$ and $\omega_H > 3000$ cm$^{-1}$. As P increases, the $\omega_H$ is softened from 3520 cm$^{-1}$ to 3320 cm$^{-1}$ and the $\omega_L$ is stiffened from 120 to 336 cm$^{-1}$, disregarding the possible phase change and other supplementary peaks at ~400 and ~3400 cm$^{-1}$. As compared in Figure 3(b), the currently MD-derived phonon relaxation is in good agreement with the results of ice-VIII measured at 80 K, using Raman[4, 23] and infrared (IR) spectroscopies.[26, 35] The consistency in the MD-derived and measured phonon relaxation dynamics of the two branches evidences further the validity of the hypothesis and the reliability of the MD derivatives.

The Raman spectroscopy could resolve the vibration of the intramolecular nonbond in the frequency range of $\omega_L < 300$ cm$^{-1}$ and that of the intermolecular bond in the frequency range of $\omega_H > 3000$ cm$^{-1}$. From the first order approximation and conventional approach for other existing materials,[36] the bonding and nonbonding part of the hydrogen bond can be taken as a harmonic system each with an interaction potential, $u_x(r)$. Equaling the vibration energy of the harmonic system to the third term of the Taylor series of its interaction potential at equilibrium, we can obtain the relation: [36]

$$\frac{1}{2}\mu(\Delta\omega)^2(r-d_x)^2 = \frac{1}{2}k_x(r-d_x)^2 \cong \frac{1}{2}\left.\frac{\partial u(r)}{\partial r^2}\right|_{r=d_x} x^2 \propto \frac{1}{2}\frac{E_x}{\mu d_x^2}(r-d_x)^2$$

$$\Delta\omega_x = \omega_x - \omega_{x0} \propto \frac{E_x^{1/2}}{\mu d_x}$$

(2)

The $k_x$ is the force constant at equilibrium and $\mu$ the reduced mass of a dimer vibrating bodies. Generally, the Raman shift depends on the length and energy of the bond and the reduced mass of the atoms or molecules of the vibronic system. From the dimensional point of view, the second order derivative of the potential at equilibrium is proportional to the bind energy $E_x$ divided by the bond length in the form of $d_x^2$. The $E_x^{1/2}/d_x \cong \sqrt{Y_x d_x}; (Y_x \approx E_x/d_x^3)$ is right the square root of the stiffness being the product of the Young's modulus and the bond length.[35] The intrinsic vibration frequency of the bond is detectable as the Raman shift from the referential point $\omega_{x0}$. This relation indicates that if the bond is compressed or the binding energy increases, a blue shift will happen, and vice versa. Therefore, the frequency shift of the soft and the stiff mode tells us the change of the length and energy of the respective partition of the hydrogen bond. The fact of the soft phonon stiffening and the stiff phonon softening supports the hypothesis that the longer-and-weaker lone-pare virtual bond becomes shorter-and-stiffer while the shorter-and-stronger real bond becomes longer-and-softer.

## 4.4  P-induced phase transition

The correlation between the bond length (*d*), bond energy (*E*), and the $T_C$ of a system under compression has been established and expressed as follows,[35]

$$\frac{T_C(P)}{T_C(P_0)} = \frac{E_x(P)}{E_x(P_0)} = 1 + \Delta_p = 1 - \frac{\int_{V_0}^{V} p(v) \mathrm{d}v}{E_x(P_0)} = 1 - \frac{\int_1^Y p(y) \mathrm{d}y}{E_x(P_0)/V_0}$$

$$= 1 - \frac{\int_{P_0}^{P} p \frac{\mathrm{d}d_x}{\mathrm{d}p} \mathrm{d}p}{E_x(P_0)/d_{x0}} = 1 - \frac{\int_{P_0}^{P} p(\alpha - 2\beta P_0 + 2\beta p) \mathrm{d}p}{E_x(P_0)/d_{x0}}$$

$P_0$ is the reference of the ambient. $E_{x,density} = E_x(P_0)/V_0$ is the binding energy density per bond with the given volume $V_0$. The known trend of $T_C$-$P$ dropping with pressure elevation justifies that the change of $T_C$ is dominated by the binding energy of the real bond, as the binding energy of the lone pair in the 1os meV level. From fitting to the measurement, we estimated the binding energy of the real bond as $E_{HO}$(1GPa) = 7.2 eV at $P_0$ = 1 GPa. Figure 4 compares the calculated $T_C$ of ice transferring from VIII to VII phase with the measurements.[23, 26, 37]

V    Conclusion

Incorporating the extended Ice Rule of Pauling and the *ab initio* MD calculations, we have been able to uncover the origin for the *P*-derived symmetric, vibronic, volumetric, and phase transitional anomalies of ice under compression. It turns out that the initially longer-and-weaker nonbond becomes shorter and slightly stiffer but the initially shorter-and-stronger real bond becomes longer and softer under increased pressure because of the resultant forces of the applied compression, the inter electron pair repulsion, and the uneven binding strength of the two segments of the hydrogen bond. The repulsion between the unevenly bounded electron pairs should form the soul dictating the physical anomalies of $H_2O$ under various conditions.

Table and figure captions

Table 1 MD derived lengths of the two parts and the resultant of the hydrogen bond of ice, with fitting parameters of $d_0$, $\alpha$, $\beta$.

| P(GPa) | O-H (Å) | O:H (Å) | O-O(Å) |
|---|---|---|---|
| 1 | 0.97358 | 1.76708 | 2.74066 |
| 5 | 0.97901 | 1.76247 | 2.74148 |
| 10 | 0.98527 | 1.75041 | 2.73568 |
| 15 | 0.99125 | 1.72133 | 2.71258 |
| 20 | 1.00245 | 1.6919 | 2.69435 |
| $d_0$ | 0.9741 | 1.7683 | 2.7415 |
| $\alpha$ ($10^{-4}$) | 9.510 | -3.477 | 1.717 |
| $\beta$ ($10^{-5}$) | 2.893 | -10.28 | -5.835 |

Figure 1 (a) the extension (2H$_2$O tetrahedron) of Pauling's "two-in two-out" ice rule (the central H$_2$O tetrahedron) allows us to focus on the cooperative interaction between the unevenly-bounded nonbonding lone pair ":" and bonding pair "–" of electrons in the quasi-linear "O$^{2-}$ : H$^{+/p}$-O$^{2-}$" hydrogen bond. (b) The proton symmetrization of ice under pressures is induced by the compressing of the nonbonding lone pair ":" and the oppositely repulsion of the bonding pair "–", moving the proton towards the center of O-O. The H$^{+/p}$ is set as the coordination origin.

Figure 2(a) MD calculation results of the pressure-induced ":" and "–" relaxation dynamics and the proton centralization occurring at 58.6 GPa at a O-O distance of 0.221 nm in exceedingly good accordance with the reported results at 59 GPa and 0.22 nm.[24] (b) the relative force constant of the bond and the nonbond indicating …. (b) Matching of the MD-derived to the measured [4] volume decay of ice.

Figure 3 (a) MD-derived power spectra of ice-VIII under pressure in comparison to (b) the measured vibration peaks of $\omega_H$(circle) and $\omega_L$(square) of ice-VIII at 80 K,[4] using Raman[23] and infrared absorption.[26] The trends consistency evidences that the $H^{+/p}:O^{2-}$ is shortened (blue shift of the low-frequency mode <400 cm$^{-1}$) and the elongated $H^{+/p}$-$O^{2-}$ (red shift of the high-frequency stretching mode >3000 cm$^{-1}$). [35]

Figure 4 Theoretical matching to the measured transition temperature of ice-VIII to proton-disordered ice-VII.[23, 26, 37]

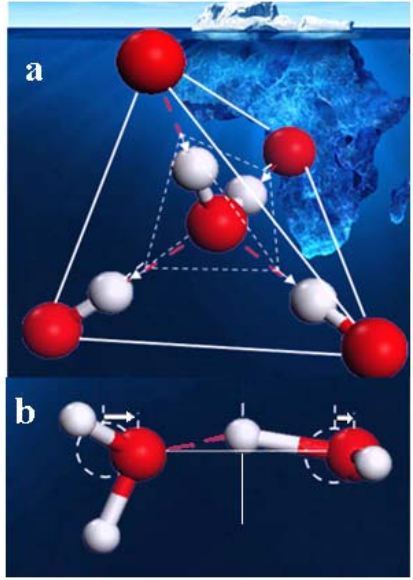

Figure 1

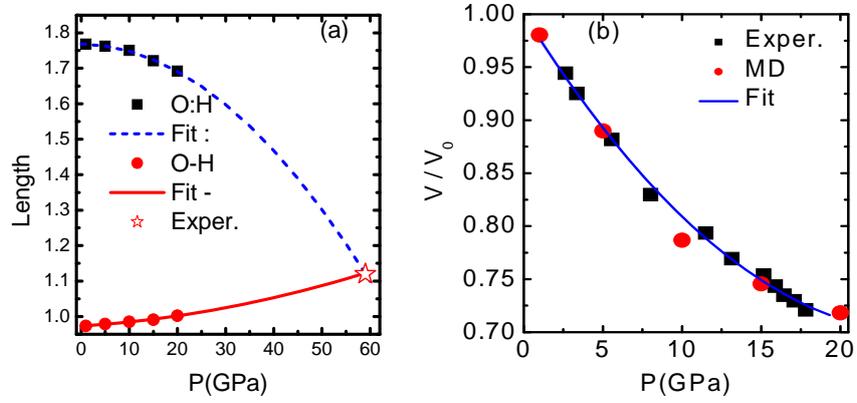

Figure 2

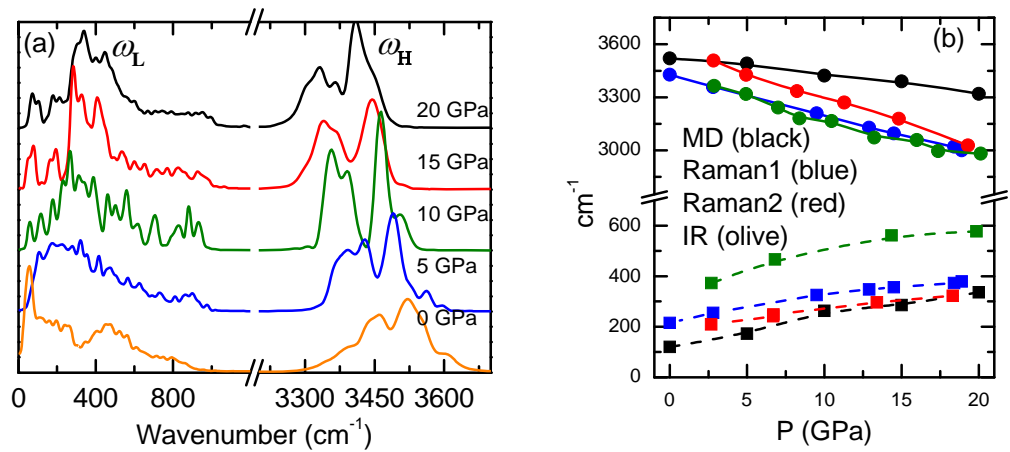

Figure 3

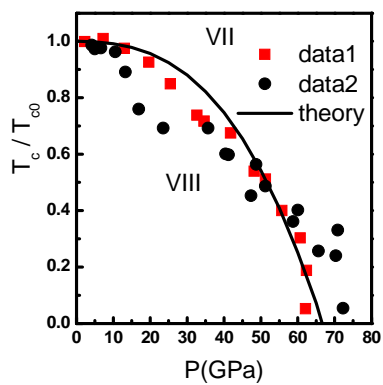

Figure 4

References

[1]  Ball P, *Water: Water - an enduring mystery.* Nature, 2008; **452**: 291-292.
[2]  Stiopkin IV, Weeraman C, Pieniazek PA, Shalhout FY, Skinner JL, and Benderskii AV, *Hydrogen bonding at the water surface revealed by isotopic dilution spectroscopy.* Nature, 2011; **474**: 192-195.
[3]  Marx D, Tuckerman ME, Hutter J, and Parrinello M, *The nature of the hydrated excess proton in water.* Nature, 1999; **397**: 601-604.
[4]  Yoshimura Y, Stewart ST, Somayazulu M, Mao H, and Hemley RJ, *High-pressure x-ray diffraction and Raman spectroscopy of ice VIII.* J. Chem. Phys., 2006; **124**: 024502.
[5]  Kang D, Dai J, Hou Y, and Yuan J, *Structure and vibrational spectra of small water clusters from first principles simulations.* J. Chem. Phys., 2010; **133**: 014302.
[6]  Smyth M and Kohanoff J, *Excess Electron Localization in Solvated DNA Bases.* Phys. Rev. Lett., 2011; **106**: 238108.
[7]  Baaske P, Duhr S, and Braun D, *Melting curve analysis in a snapshot.* Appl. Phys. Lett., 2007; **91**: 133901.

[8] Castellano C, Generosi J, Congiu A, and Cantelli R, *Glass transition temperature of water confined in lipid membranes as determined by anelastic spectroscopy.* Appl. Phys. Lett., 2006; **89**: 233905.

[9] Park JH and Aluru NR, *Water film thickness-dependent conformation and diffusion of single-strand DNA on poly(ethylene glycol)-silane surface.* Appl. Phys. Lett., 2010; **96**: 123703.

[10] Kobayashi K, Koshino M, and Suenaga K, *Atomically Resolved Images of I(h) Ice Single Crystals in the Solid Phase.* Phys. Rev. Lett., 2011; **106**: 206101.

[11] Hermann A and Schwerdtfeger P, *Blueshifting the Onset of Optical UV Absorption for Water under Pressure.* Phys. Rev. Lett., 2011; **106**: 187403.

[12] Chen W, Wu XF, and Car R, *X-Ray Absorption Signatures of the Molecular Environment in Water and Ice.* Phys. Rev. Lett., 2010; **105**: 017802.

[13] Lin CK, Wu CC, Wang YS, Lee YT, Chang HC, Kuo JL, and Klein ML, *Vibrational predissociation spectra and hydrogen-bond topologies of H+ (H2O)(9-11).* PCCP, 2005; **7**: 938-944.

[14] Lenz A and Ojamae L, *A theoretical study of water equilibria: The cluster distribution versus temperature and pressure for (H2O)(n), n=1-60, and ice.* Journal of Chemical Physics, 2009; **131**.

[15] Frenken JWM and Oosterkamp TH, *MICROSCOPY When mica and water meet.* Nature, 2010; **464**: 38-39.

[16] Headrick JM, Diken EG, Walters RS, Hammer NI, Christie RA, Cui J, Myshakin EM, Duncan MA, Johnson MA, and Jordan KD, *Spectral signatures of hydrated proton vibrations in water clusters.* Science, 2005; **308**: 1765-1769.

[17] Gregory JK, Clary DC, Liu K, Brown MG, and Saykally RJ, *The water dipole moment in water clusters.* Science, 1997; **275**: 814-817.

[18] Bjerrum N, *Structure and properties of ice.* Science, 1952; **115**: 385-390.

[19] Soper AK, Teixeira J, and Head-Gordon T, *Is ambient water inhomogeneous on the nanometer-length scale?* Proceedings of the National Academy of Sciences of the United States of America, 2010; **107**: E44-E44.

[20] Zha C-S, Hemley RJ, Gramsch SA, Mao H-k, and Bassett WA, *Optical study of H2O ice to 120GPa: Dielectric function, molecular polarizability, and equation of state* J. Chem. Phys., 2007; **126**: 074506.

[21] Sun CQ, *Thermo-mechanical behavior of low-dimensional systems: The local bond average approach.* Prog. Mater Sci., 2009; **54**: 179-307.

[22] Errandonea D, Schwager B, Ditz R, Gessmann C, Boehler R, and Ross M, *Systematics of transition-metal melting.* Phys Rev B, 2001; **63**: 132104.

[23] Pruzan P, Chervin JC, Wolanin E, Canny B, Gauthier M, and Hanfland M, *Phase diagram of ice in the VII–VIII–X domain. Vibrational and structural data for stronglycompressed ice VIII.* J. Raman Spectrosc., 2003; **34**: 591-610.

[24] Benoit M, Marx D, and Parrinello M, *Tunnelling and zero-point motion in high-pressure ice.* Nature, 1998; **392**: 258-261.

[25] Zheng WT and Sun CQ, *Underneath the fascinations of carbon nanotubes and graphene nanoribbons.* Energy & Environmental Science, 2011; **4**: 627-655.

[26] Song M, Yamawaki H, Fujihisa H, Sakashita M, and Aoki K, *Infrared absorption study of Fermi resonance and hydrogen-bond symmetrization of ice up to 141 GPa.* Physical Review B, 1999; **60**: 12644.

[27] Pauling L, *The structure and entropy of ice and of other crystals with some randomness of*

[27] *atomic arrangement.* J. Am. Chem. Soc., 1935; **57**: 2680-2684.
[28] Sun CQ, *Dominance of broken bonds and nonbonding electrons at the nanoscale.* Nanoscale, 2010; **2**: 1930-1961.
[29] Sun CQ, *Oxidation electronics: bond-band-barrier correlation and its applications.* Prog. Mater Sci., 2003; **48**: 521-685.
[30] Sun H, *COMPASS: An ab Initio Forcefield Optimized for Condensed-Phase Applications - Overview with Details on Alkane and Benzene Compounds.* J. Phys. Chem. B, 1998; **102**: 7338.
[31] Wang J, Qin QH, Kang YL, Li XQ, and Rong QQ, *Viscoelastic adhesive interfacial model and experimental characterization for interfacial parameters.* Mech. Mater., 2010; **42**: 537-547.
[32] Liu K, Cruzan JD, and Saykally RJ, *Water clusters.* Science, 1996; **271**: 929-933.
[33] Ludwig R, *Water: From clusters to the bulk.* Angewandte Chemie-International Edition, 2001; **40**: 1808-1827.
[34] Holzapfel W, *On the Symmetry of the Hydrogen Bonds in Ice VII.* J. Chem. Phys., 1972; **56**: 712.
[35] Sun CQ, *Size dependence of nanostructures: Impact of bond order deficiency.* Prog. Solid State Chem., 2007; **35**: 1-159.
[36] Yang XX, Li JW, Zhou ZF, Wang Y, Zheng WT, and Sun CQ, *Frequency response of graphene phonons to heating and compression.* Appl. Phys. Lett., 2011; **Accepted on 13/09**.
[37] Aoki K, Yamawaki H, and Sakashita M, *Observation of Fano Interference in High-Pressure Ice VII.* Physical Review Letters, 1996; **76**: 784.